\documentclass{nature}

\usepackage{amsmath}
\usepackage{amssymb}
\usepackage{graphicx}

\title{Recurrent infomax generates cell assemblies, avalanches, and simple cell-like selectivity}
\author{Takuma Tanaka$^1$, Takeshi Kaneko$^{1,2}$ \& Toshio Aoyagi $^{2,3}$}
\date{}
\begin{document}
\maketitle

\begin{affiliations}
\item Department of Morphological Brain Science, Graduate School
of Medicine, Kyoto University, Japan
\item CREST, JST
\item Department of Applied Analysis and Complex Dynamical Systems,
Graduate School of Informatics, Kyoto University, Japan
\end{affiliations}

\begin{abstract}
Through evolution, animals have acquired central nervous systems (CNSs),
 which are extremely efficient
 information processing devices that improve an animal's adaptability to various
 environments.
It has been proposed that the process of information
 maximization (infomax\cite{Linsker1988}), which maximizes the information transmission from
 the input to the output of a feedforward network, may provide an
 explanation of
 the stimulus selectivity of neurons in CNSs\cite{Tsukada1975,Atick1992,BellSejnowski1995:NeuralComput:1129-1159,OlshausenField1996:Nature:607-609,Bell1997,Lewicki2002}.
However, CNSs contain not only feedforward but also recurrent synaptic connections,
 and little is known about
 information retention over time in such recurrent networks.
Here, we propose a learning algorithm based on infomax
 in a recurrent network, which we call ``recurrent infomax'' (RI).
RI maximizes information retention and thereby
 minimizes information loss in a network.
We find that feeding in external inputs consisting of information
 obtained from photographs of natural scenes into an
 RI-based model of a recurrent network
 results in the appearance of Gabor-like selectivity quite similar to
that existing in simple cells of the primary visual cortex (V1).
More importantly, we find that without external input, this network exhibits
cell assembly-like and synfire chain-like
 spontaneous activity\cite{Hebb1949,Abeles1991,Diesmann1999} and a critical
 neuronal avalanche\cite{Beggs2003,Teramae2007,Abbott2007}.
RI provides a simple framework to explain a wide range of phenomena
observed in \textit{in vivo} and
\textit{in vitro} neuronal networks,
and it should provide  a novel understanding of experimental results for
 multineuronal activity and plasticity
 from an information-theoretic point of view.
\end{abstract}
%\end{document}

%\begin{document}

Recent advances in multineuronal
recording have allowed us to observe phenomena in
the recurrent networks of CNSs that are much more complex than
previously thought to exist.
The existence of interesting type of neuronal activity, such as
patterned firing, synchronization, oscillation, and global state
transitions has been
revealed by multielectrode recording and calcium imaging\cite{Nadasdy1999,Cossart2003,Ikegaya2004,Fujisawa2006,Sakurai2006}.
However, in contrast to the rapidly accumulating experimental data, 
theoretical works attempting to account for this wide range of data are developing more slowly.
To understand the behaviour exhibited by recurrent neuronal networks of CNSs, we
investigated a network employing an
RI algorithm that maximizes information retention.
The role of RI is to allow a recurrent network
to optimize the synaptic connection weight in order
to maximize information retention and
thereby minimize information loss
by maximizing the mutual information of the temporally
successive states of the network.

Here we briefly describe our recurrent network model,
leaving the details to the Supplementary Notes.
In this model, neurons were connected according to the weight matrix $W_{ij}$, and
their firing states [$x_i(t)=1$ (fire) and 0 (not fire)]
at time step $t$ are synchronously updated to time step $t+1$.
The firing state $x_i(t+1)$ of neuron $i$ at time step $t+1$ is
determined stochastically with the firing probability
\begin{equation}
 p_i(t+1)=\frac{p_{\mathrm{max}}}{1+\exp\left(-\sum_j W_{i
				    j}(x_j(t)-\bar{p}_j)+h_i(t)\right)}, \label{probability}
\end{equation}
where $h_i(t)$ is the threshold of neuron $i$, which is adjusted to
fix the mean firing probability of
neuron $j$ to $\bar{p}_j$, and $p_\mathrm{max}$ is the maximal firing probability.
When $p_\mathrm{max}=0.5$, a neuron fires, on
average, only once, even if the neuron receives a sufficiently strong excitatory input twice.
A small value of $p_\mathrm{max}$ thus makes the firing of the neurons quite unreliable.
%In contrast, if $p_\mathrm{max}$ is close to 1, 
%it is highly probable that a strong input makes a neuron fire.
Thus, $p_\mathrm{max}$ determines the reliability with which a model
neuron fires in response to an input.

\begin{figure}
\begin{center}
\includegraphics[width=183mm,bb=0 0 1047 1041]{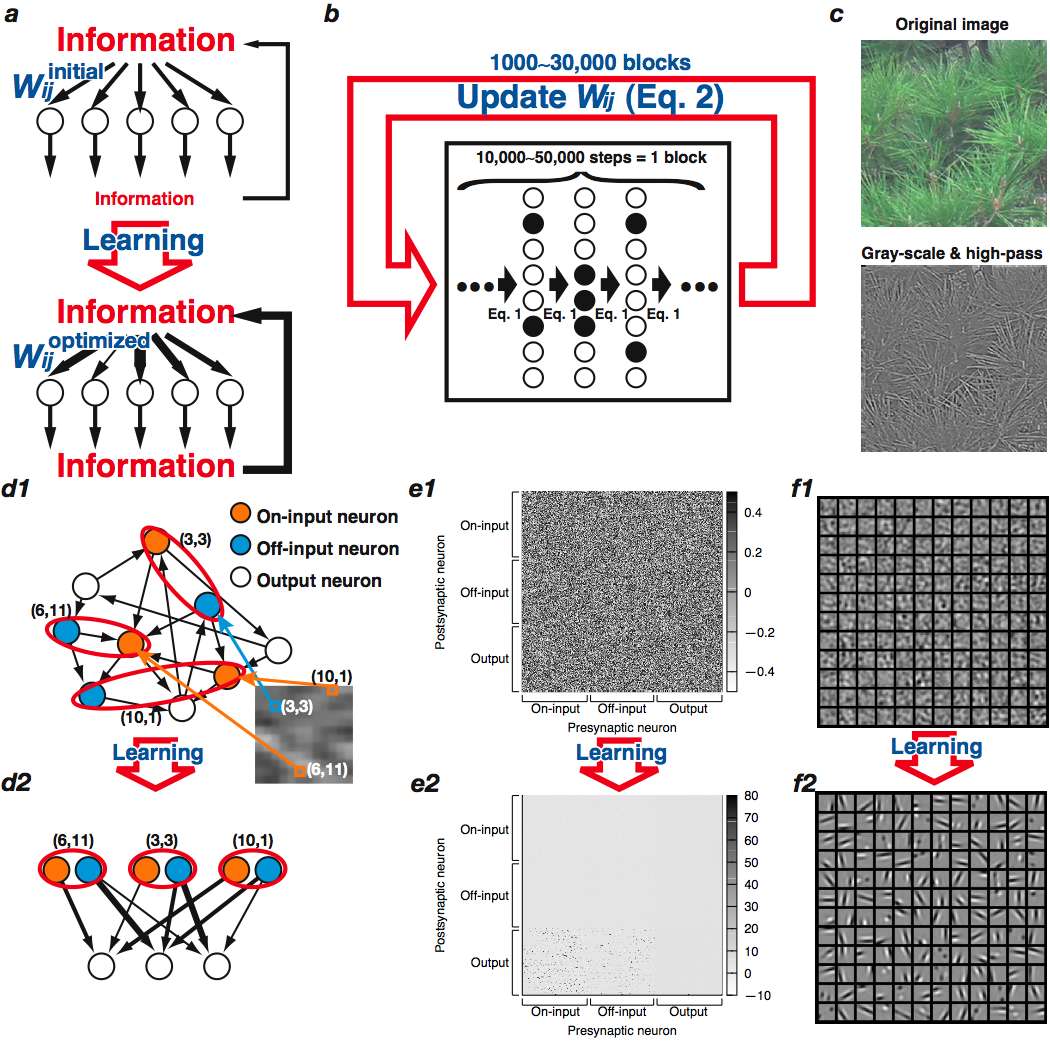}\\Figure 1%0 0 1047 1041 502.56 503.04
\caption{\label{natural}
Formation of the feedforward structure through an algorithm based on RI in the model network with
 external input.
(a) Learning changes the initial weight matrix
 $W_{ij}^\mathrm{initial}$ to $W_{ij}^\mathrm{optimized}$ so as to
 maximize the information retained in a recurrent network.
(b)
In the simulation, $W_{ij}$ was updated using Eq.~\ref{update} at the end
 of each block, which consists of 10,000-50,000 steps.
(c)
The original photograph ($1024\times 1024$) of a pine tree was converted to a
 gray-scaled, high-pass filtered image.
Image patches ($12\times 12$) randomly selected from the high-pass
 filtered image were used as the external inputs to the network at each time step.
(d1,e1)
Initially, 432 neurons were connected according to a random weight
 matrix.
Of these neurons
 144 were on-input, 144 were off-input, and 144 were output neurons.
Each of the 144 dots in an image patch was linked to a pair of an on- and
an off-input neuron in such a manner that the on-input and off-input neurons were set to 1
 (fire) only when the corresponding dot had a positive and negative sign,
 respectively.
Output neurons fired spontaneously according to Eq.~\ref{probability}.
(d2,e2)
After learning, feedforward structure from
 input to output neurons appeared in the model network.
(f1,2)
Averaging the image patches that evoked firings of the output neurons
revealed that the output neurons, which did not exhibit clear selectivity
 before learning,
 responded to the Gabor-like stimulus after learning.
}
\end{center}
\end{figure}

To maximize information retention,
our recurrent network starts from a random
weight $W_{ij}^\mathrm{initial}$ and develops toward an optimized network with
$W_{ij}^\mathrm{optimized}$ (Fig.~\ref{natural}a), in a manner
determined by
the gradient ascent algorithm,
\begin{eqnarray}
W_{ij}\leftarrow W_{ij}+\eta \frac{\partial I}{\partial W_{ij}},
 \label{update}
\end{eqnarray}
where $I$ is the mutual information of two successive
states of the network and $\eta$ is the learning rate.
We performed a block of simulation consisting of 10,000-50,000 time steps,
updated $W_{ij}$ at the end of the block, and then started the calculation for
the next block (Fig.~\ref{natural}b).

We first observed the behaviour of this model network under external input.
Image patches from a photograph preprocessed by a high-pass filter
were used as the external input (Fig.~\ref{natural}c).
The neurons in this network were divided into three groups.
144 on-input
and 144 off-input neurons, and the 144 output neurons were randomly selected
from the network (Fig.~\ref{natural}d1).
Dots with positive and negative values in a randomly selected
$12\times 12$ image patch
excited the corresponding on-input and off-input neurons, respectively.
The states of the input neurons were stochastically set to 1 or 0 with
firing probabilities proportional to the intensities of the corresponding dots,
whereas 
the states of the output neurons were not set by the external input.
Instead, the firings of these
neurons were determined by Eq.~\ref{probability} with 
$p_\mathrm{max}=0.95$.
Initially, the connection weight $W_{ij}$ was a random matrix (Fig.~\ref{natural}e1),
and therefore the output neurons did not exhibit clear selectivity with
respect to the
external input from the input neurons (Fig.~\ref{natural}f1).
After learning, however, the network self-organized a feedforward structure from
the on-input and off-input neurons to the output neurons (Fig.~\ref{natural}e2,d2).
Averaging the image patches that evoked firings in an output neuron,
we found that the output neuron became
highly selective to Gabor function-like stimuli (Fig.~\ref{natural}f2),
exhibiting behaviour quite similar to the selectivity of simple cells in
the V1 cortex\cite{HubelWiesel1959}.
Our optimization algorithm based on RI hence caused the model
network to become organized into a feedforward network containing simple cell-like output neurons.

\begin{figure}
\begin{center}
\includegraphics[width=183mm,bb=0 0 1093 894]{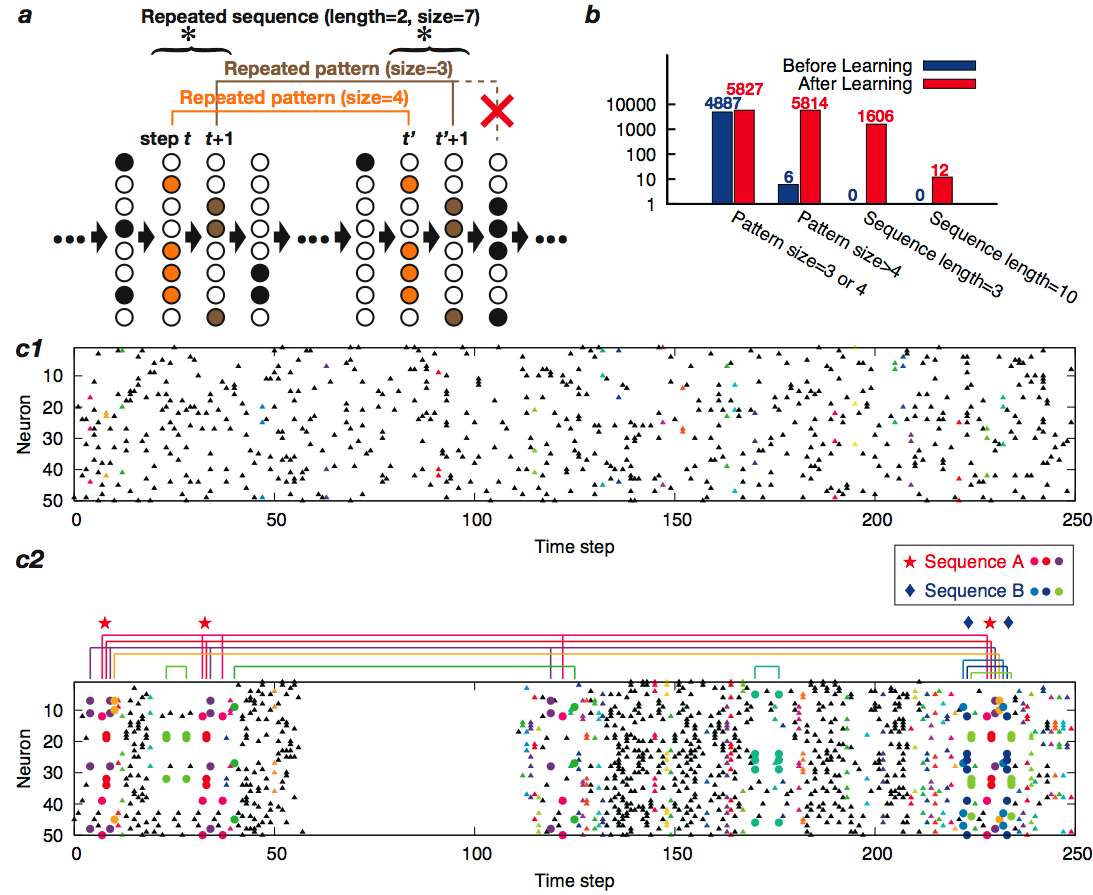}\\Figure 2 %1093 894 524.64 429.12
\caption{\label{sequence}
Repeated spatial patterns and spatiotemporal sequences occurred frequently
 in the network with $p_\mathrm{max}=0.95$ after learning.
(a)
We define a repeated pattern as a spatial firing pattern that is identically repeated at
 different time steps.
The size of a pattern is defined as the number of neurons firing in the pattern.
A sequence that contains a particular set of patterns appearing repeatedly in
 the same temporal order
 is called a ``repeated sequence.'' The size of a repeated sequence is
 defined as the sum of the sizes of the
 patterns contained in it.
(b)
The numbers of occurrences of the patterns and sequences repeated 
in the test block (50,000 steps) were compared before and after learning.
In this histogram, only the sequences with sizes larger than $5l$,
 where $l$ is the length of the sequence, were counted.
(c1,2)
When the repeated patterns in the 50,000 steps were coloured, 
it was found that no pattern occurred more than once in this short raster plot before learning (c1).
By contrast, several patterns appeared multiple times in the raster plot
 after learning (c2).
In addition, repeated sequences were found only in the raster plot after
 learning (red stars and blue diamonds).
}
\end{center}
\end{figure}

In the simulation described above, the external input was fed into the network with high response reliability ($p_\mathrm{max}=0.95$).
Next, we examined the evolution of the spontaneous activity in the
neuronal network without external input.
To identify repeated activity in the model network,
we defined a repeated pattern as a spatial pattern of
neuronal firings that occurs at least twice in a test
block (Fig.~\ref{sequence}a).
Colouring repeated patterns consisting of $\ge 3$ firing neurons in
raster plots of the network (Fig.~\ref{sequence}c1,c2), we found that
the number of
repeated patterns increased after learning.
Several patterns were repeated in a sample of 250 steps
as seen in Fig.~\ref{sequence}c2, where
the repeated patterns are indicated by consistently coloured circles and connected by lines.
Moreover, some patterns appeared to constitute repeated sequences.
For example, sequence A, composed of the magenta, orange, and purple
patterns, appears three times in Fig.~\ref{sequence}c2.
To be more quantitative, we tabulated the numbers of
occurrences of repeated patterns and sequences, and compared these numbers before and after
learning (Fig.~\ref{sequence}b).
We found that both repeated patterns and repeated sequences increased significantly
after learning.
This indicates that the present algorithm embeds not only repeated
patterns but also repeated sequences of firings
into the network structure as a result of the optimization.
Thus, when a pattern in a sequence is activated at
one step, it is highly probable that the next pattern in that sequence
will be activated at the next step.
This predictability means that the state of the network at one time step shares much
information with the state at the next time step.
Hence, we concluded that the repeated activation of an embedded sequence is an efficient
way to maximize information retention in a recurrent network.
These repeated patterns and sequences have been experimentally observed
\textit{in vivo}\cite{Skaggs1996,Sakurai2006,Yao2007} and \textit{in
vitro}\cite{Cossart2003,Ikegaya2004}, and their existence is suggested by the
theory of cell assemblies proposed by Hebb\cite{Hebb1949} and the theory
of synfire chains proposed  by
Abeles\cite{Abeles1991}.
We thus see that RI accounts for the appearance of cell assemblies, sequences, and
synfire chains in neuronal networks.

\begin{figure}
\begin{center}
\includegraphics[width=183mm,bb=0 0 1073 472]{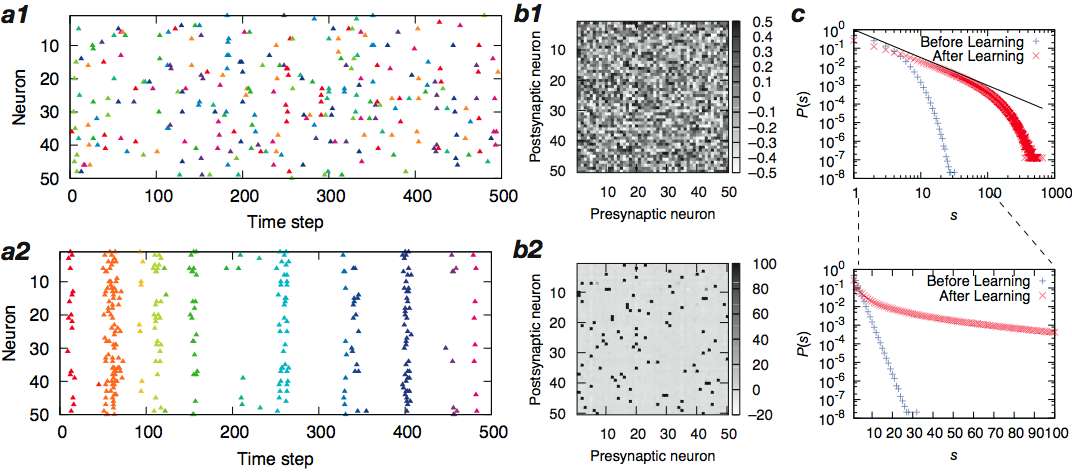}\\Figure 3 %1073 472 515.04 226.56
\caption{\label{avalanche}
Spontaneous activity of the recurrent network with $p_\mathrm{max}=0.5$.
(a1,2) Individual bursts in the spontaneous activity before (a1) and
 after learning (a2) are indicated by different colours.
The bursts before learning were short and frequently interrupted by
 steps without firing, whereas the bursts after learning had much
 longer durations.
(b1,2) The initial $W_{ij}$ with random weights evolved into a matrix
 with relatively few strong weights.
Most rows and columns contained two strong excitatory connections (black dots); that is,
 most neurons had two strong inputs and two strong outputs.
(c) %Double-log (above) and mono-log (below) plot of the size
Frequency distribution $P(s)$ of the burst size plotted as a function of
 the size, $s$.
The black line corresponds to a slope of $-1.5$.
}
\end{center}
\end{figure}

We next examined the behaviour of the same spontaneous model in the case
that the
maximal firing probability was small ($p_\mathrm{max}=0.5$).
For small $p_\mathrm{max}$, the number of identically repeated
sequences is small,
and the network seems to lose structured activity.
However, we found characteristic network activity consisting of firing bursts (Fig.~\ref{avalanche}a2),
which are defined as consecutive firing steps that are immediately  preceded and
followed by ``silent'' steps, with no firing.
We found that after learning,
the distribution $P(s)$ of the burst size, $s$, the total number of
firings in a burst, obeys a power-law distribution
$P(s)\propto s^{\gamma}$ with $\gamma\approx -1.5$, whereas,
before learning, we have
$P(s)\propto\exp(-\alpha s)$ (Fig.~\ref{avalanche}c).
This result is consistent with experimental results.
Beggs and Plenz\cite{Beggs2003} recorded the spontaneous activity of an
organotypic culture from a cortex using multielectrode arrays.
Defining an avalanche similarly to our burst, they
found that the size distribution of avalanches is accurately fit by
a power-law distribution with exponent $-1.5$\cite{Beggs2003}.
To explain this, they argued that
a neuronal network is tuned to minimize the information loss and
that this is realized when one firing induces an
average of one firing at the next step.
They showed that this condition yields the universal exponent
$-3/2$, using the self-organized criticality of the sandpile
model\cite{Bak1987,Harris1989}.
This condition also holds for the present network, because 
each neuron with $p_\mathrm{max}=0.5$ after learning
had two strong input connections and two strong output
connections  on average (Fig.~\ref{avalanche}b2).
The universal exponent $-3/2$ was observed in the network for small
$p_\mathrm{max}$, but for $p_\mathrm{max}=0.95$, the size distribution of
bursts $P(s)$ in the system did not exhibit a power-law distribution, and displayed
 several peaks, reflecting the existence of stereotyped sequences (see
 Supplementary Notes).
We thus conclude that RI embeds information-efficient structures
in which one firing induces on average one firing at the
next step in a network with small $p_\mathrm{max}$.

\begin{figure}
\begin{center}
\includegraphics[width=183mm,bb=0 0 1062 460]{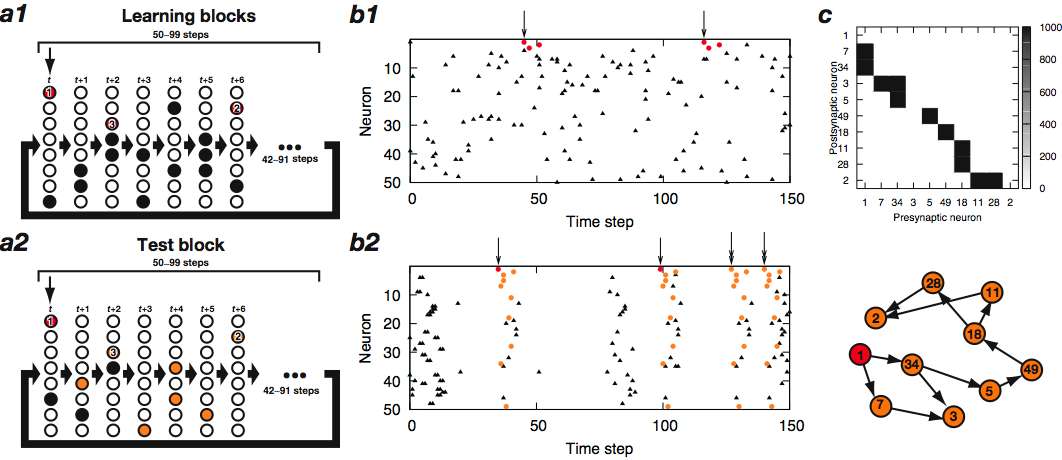}\\Figure 4 %1062 460 509.76 220.8
\caption{\label{trigger}
A feedforward structure was embedded in
 the model network by the temporally-structured stimulation.
(a1,2)
In the learning blocks, the state of neuron 1 was set to 1 (fire) at
 random intervals ranging from 50 to 99
 steps. The first time step, $t$, is indicated by the arrow in a1.
At $t+2$, the state of neuron 3 was set to
 1, and at $t+6$, the state of  neuron 2 was set to 1.
In the test block after learning, only neuron 1 was set to 1 at
random intervals ranging from 50 to 99 steps (a2).
External stimulations are indicated by red circles.
(b1, 2) The network activity in an early learning block (b1) and the test block (b2).
The steps at which neuron 1 was set to 1 are indicated by arrows, and
 externally evoked firings of neurons 1, 2, and 3 are indicated by red circles.
Although the states of neurons 2 and 3 were not set from the outside during
 the test block, neurons 2 and 3 fired spontaneously six and two steps, respectively,
 after neuron 1 fired (as indicated by orange circles).
The sequence of firings embedded by learning was replayed after the
 spontaneous firing of neuron 1 (double arrows).
(c)
The weight matrix of the network after learning (top) and its schematic
 representation (bottom) indicate a feedforward structure which
 underlies the firing sequence starting from neuron 1 and containing neurons 3
 and 2.
}
\end{center}
\end{figure}

To reveal the essential mechanism responsible for the behaviour
described above, we returned to the recurrent network with
an external input (Fig.~\ref{trigger}).
In the learning blocks, we repeatedly stimulated neurons 1, 3, and 2 in
sequence (Fig.~\ref{trigger}a1,b1).
In the successive test block, in which only neuron 1 was stimulated
externally (Fig.~\ref{trigger}a2), the firing of neuron 1 was followed by
spontaneous firings of neurons 3 and 2 (Fig.~\ref{trigger}b2, arrows),
because, as we saw above, embedding a sequence of firings into the
network structure is an efficient way to retain information.
In addition, the spontaneous firing of neuron 1 triggers the
sequence containing the firings of neurons 3 and 2
(Fig.~\ref{trigger}b2, double arrows).
The form of the weight matrix after learning reveals that
a feedforward structure starting from neuron 1 
was embedded in the network (Fig.~\ref{trigger}c).
It is thus seen that
RI embeds externally input temporal firing patterns into the network by
producing feedforward structures, and, as a result, the network can spontaneously
reproduce the patterns.

In this study we have found that RI acts to optimize the network
structure by maximizing the information retained in the recurrent network.
Simple cell-like activity, repeated sequences, and
neuronal avalanches were realized in the model network.
These characteristic types of activity resulted from the
network structure embedded by the optimization algorithm.
On the basis of these results, we conjecture that RI underlies the neuronal plasticity rule 
generating these structures and activity.
We believe that RI will help us to understand the meaning of
\textit{in vivo} and \textit{in vitro} experimental results,
particularly to characterize
the spontaneous activity of neurons in the context of
information theory.
Our next goal is to derive a plasticity rule in a bottom-up way
employing RI, and to compare this rule with
experimentally obtained plasticity rules.
%\end{document}
\bibliographystyle{naturemag}
\bibliography{paper}

\begin{thebibliography}{10}
\expandafter\ifx\csname url\endcsname\relax
  \def\url#1{\texttt{#1}}\fi
\expandafter\ifx\csname urlprefix\endcsname\relax\def\urlprefix{URL }\fi
\providecommand{\bibinfo}[2]{#2}
\providecommand{\eprint}[2][]{\url{#2}}

\bibitem{Linsker1988}
\bibinfo{author}{Linsker, R.}
\newblock \bibinfo{title}{Self-organization in a perceptual network}.
\newblock \emph{\bibinfo{journal}{Computer}} \textbf{\bibinfo{volume}{21}},
  \bibinfo{pages}{105--117} (\bibinfo{year}{1988}).

\bibitem{Tsukada1975}
\bibinfo{author}{Tsukada, M.}, \bibinfo{author}{Ishii, N.} \&
  \bibinfo{author}{Sato, R.}
\newblock \bibinfo{title}{{{T}emporal pattern discrimination of impulse
  sequences in the computer-simulated nerve cells}}.
\newblock \emph{\bibinfo{journal}{Biol.\ Cybern.}}
  \textbf{\bibinfo{volume}{17}}, \bibinfo{pages}{19--28}
  (\bibinfo{year}{1975}).

\bibitem{Atick1992}
\bibinfo{author}{Atick, J.~J.}
\newblock \bibinfo{title}{{Could information theory provide an ecological
  theory of sensory processing?}}
\newblock \emph{\bibinfo{journal}{Network}} \textbf{\bibinfo{volume}{3}},
  \bibinfo{pages}{213--251} (\bibinfo{year}{1992}).

\bibitem{BellSejnowski1995:NeuralComput:1129-1159}
\bibinfo{author}{Bell, A.~J.} \& \bibinfo{author}{Sejnowski, T.~J.}
\newblock \bibinfo{title}{{An information-maximization approach to blind
  separation and blind deconvolution.}}
\newblock \emph{\bibinfo{journal}{Neural\ Comput.}}
  \textbf{\bibinfo{volume}{7}}, \bibinfo{pages}{1129--1159}
  (\bibinfo{year}{1995}).

\bibitem{OlshausenField1996:Nature:607-609}
\bibinfo{author}{Olshausen, B.~A.} \& \bibinfo{author}{Field, D.~J.}
\newblock \bibinfo{title}{{Emergence of simple-cell receptive field properties
  by learning a sparse code for natural images.}}
\newblock \emph{\bibinfo{journal}{Nature}} \textbf{\bibinfo{volume}{381}},
  \bibinfo{pages}{607--609} (\bibinfo{year}{1996}).

\bibitem{Bell1997}
\bibinfo{author}{Bell, A.~J.} \& \bibinfo{author}{Sejnowski, T.~J.}
\newblock \bibinfo{title}{The `independent components' of natural scenes are
  edge filters}.
\newblock \emph{\bibinfo{journal}{Vision\ Res.}} \textbf{\bibinfo{volume}{37}},
  \bibinfo{pages}{3327--3338} (\bibinfo{year}{1997}).

\bibitem{Lewicki2002}
\bibinfo{author}{Lewicki, M.~S.}
\newblock \bibinfo{title}{{{E}fficient coding of natural sounds}}.
\newblock \emph{\bibinfo{journal}{Nat.\ Neurosci.}}
  \textbf{\bibinfo{volume}{5}}, \bibinfo{pages}{356--363}
  (\bibinfo{year}{2002}).

\bibitem{Hebb1949}
\bibinfo{author}{Hebb, D.~O.}
\newblock \emph{\bibinfo{title}{{The Organization of Behavior; a
  Neuropsychological Theory}}} (\bibinfo{publisher}{Wiley},
  \bibinfo{address}{New York}, \bibinfo{year}{1949}).

\bibitem{Abeles1991}
\bibinfo{author}{Abeles, M.}
\newblock \emph{\bibinfo{title}{{Corticonics}}} (\bibinfo{publisher}{Cambridge.
  Univ. Press}, \bibinfo{address}{Cambridge}, \bibinfo{year}{1991}).

\bibitem{Diesmann1999}
\bibinfo{author}{Diesmann, M.}, \bibinfo{author}{Gewaltig, M.~O.} \&
  \bibinfo{author}{Aertsen, A.}
\newblock \bibinfo{title}{{{S}table propagation of synchronous spiking in
  cortical neural networks}}.
\newblock \emph{\bibinfo{journal}{Nature}} \textbf{\bibinfo{volume}{402}},
  \bibinfo{pages}{529--533} (\bibinfo{year}{1999}).

\bibitem{Beggs2003}
\bibinfo{author}{Beggs, J.~M.} \& \bibinfo{author}{Plenz, D.}
\newblock \bibinfo{title}{{{N}euronal avalanches in neocortical circuits}}.
\newblock \emph{\bibinfo{journal}{J.\ Neurosci.}}
  \textbf{\bibinfo{volume}{23}}, \bibinfo{pages}{11167--11177}
  (\bibinfo{year}{2003}).

\bibitem{Teramae2007}
\bibinfo{author}{Teramae, J.-N.} \& \bibinfo{author}{Fukai, T.}
\newblock \bibinfo{title}{{{L}ocal cortical circuit model inferred from
  power-law distributed neuronal avalanches}}.
\newblock \emph{\bibinfo{journal}{J.\ Comput.\ Neurosci.}}
  \textbf{\bibinfo{volume}{22}}, \bibinfo{pages}{301--312}
  (\bibinfo{year}{2007}).

\bibitem{Abbott2007}
\bibinfo{author}{Abbott, L.} \& \bibinfo{author}{Rohrkemper, R.}
\newblock \bibinfo{title}{{{A} simple growth model constructs critical
  avalanche networks}}.
\newblock \emph{\bibinfo{journal}{Prog. Brain Res.}}
  \textbf{\bibinfo{volume}{165}}, \bibinfo{pages}{13--19}
  (\bibinfo{year}{2007}).

\bibitem{Nadasdy1999}
\bibinfo{author}{Nadasdy, Z.}, \bibinfo{author}{Hirase, H.},
  \bibinfo{author}{Czurko, A.}, \bibinfo{author}{Csicsvari, J.} \&
  \bibinfo{author}{Buzsaki, G.}
\newblock \bibinfo{title}{{{R}eplay and time compression of recurring spike
  sequences in the hippocampus}}.
\newblock \emph{\bibinfo{journal}{J.\ Neurosci.}}
  \textbf{\bibinfo{volume}{19}}, \bibinfo{pages}{9497--9507}
  (\bibinfo{year}{1999}).

\bibitem{Cossart2003}
\bibinfo{author}{Cossart, R.}, \bibinfo{author}{Aronov, D.} \&
  \bibinfo{author}{Yuste, R.}
\newblock \bibinfo{title}{{{A}ttractor dynamics of network {U}{P} states in the
  neocortex}}.
\newblock \emph{\bibinfo{journal}{Nature}} \textbf{\bibinfo{volume}{423}},
  \bibinfo{pages}{283--288} (\bibinfo{year}{2003}).

\bibitem{Ikegaya2004}
\bibinfo{author}{Ikegaya, Y.} \emph{et~al.}
\newblock \bibinfo{title}{{{S}ynfire chains and cortical songs: temporal
  modules of cortical activity}}.
\newblock \emph{\bibinfo{journal}{Science}} \textbf{\bibinfo{volume}{304}},
  \bibinfo{pages}{559--564} (\bibinfo{year}{2004}).

\bibitem{Fujisawa2006}
\bibinfo{author}{Fujisawa, S.}, \bibinfo{author}{Matsuki, N.} \&
  \bibinfo{author}{Ikegaya, Y.}
\newblock \bibinfo{title}{{{S}ingle neurons can induce phase transitions of
  cortical recurrent networks with multiple internal {S}tates}}.
\newblock \emph{\bibinfo{journal}{Cereb.\ Cortex}}
  \textbf{\bibinfo{volume}{16}}, \bibinfo{pages}{639--654}
  (\bibinfo{year}{2006}).

\bibitem{Sakurai2006}
\bibinfo{author}{Sakurai, Y.} \& \bibinfo{author}{Takahashi, S.}
\newblock \bibinfo{title}{{{D}ynamic synchrony of firing in the monkey
  prefrontal cortex during working-memory tasks}}.
\newblock \emph{\bibinfo{journal}{J.\ Neurosci.}}
  \textbf{\bibinfo{volume}{26}}, \bibinfo{pages}{10141--10153}
  (\bibinfo{year}{2006}).

\bibitem{HubelWiesel1959}
\bibinfo{author}{Hubel, D.~H.} \& \bibinfo{author}{Wiesel, T.~N.}
\newblock \bibinfo{title}{{Receptive fields of single neurones in the cat's
  striate cortex}}.
\newblock \emph{\bibinfo{journal}{J.\ Physiol.}}
  \textbf{\bibinfo{volume}{148}}, \bibinfo{pages}{574--591}
  (\bibinfo{year}{1959}).

\bibitem{Skaggs1996}
\bibinfo{author}{Skaggs, W.~E.} \& \bibinfo{author}{McNaughton, B.~L.}
\newblock \bibinfo{title}{{{R}eplay of neuronal firing sequences in rat
  hippocampus during sleep following spatial experience}}.
\newblock \emph{\bibinfo{journal}{Science}} \textbf{\bibinfo{volume}{271}},
  \bibinfo{pages}{1870--1873} (\bibinfo{year}{1996}).

\bibitem{Yao2007}
\bibinfo{author}{Yao, H.}, \bibinfo{author}{Shi, L.}, \bibinfo{author}{Han,
  F.}, \bibinfo{author}{Gao, H.} \& \bibinfo{author}{Dan, Y.}
\newblock \bibinfo{title}{{{R}apid learning in cortical coding of visual
  scenes}}.
\newblock \emph{\bibinfo{journal}{Nat.\ Neurosci.}}
  \textbf{\bibinfo{volume}{10}}, \bibinfo{pages}{772--778}
  (\bibinfo{year}{2007}).

\bibitem{Bak1987}
\bibinfo{author}{Bak, P.}, \bibinfo{author}{Tang, C.} \&
  \bibinfo{author}{Wiesenfeld, K.}
\newblock \bibinfo{title}{Self-organized criticality: An explanation of the 1/f
  noise}.
\newblock \emph{\bibinfo{journal}{Phys.\ Rev.\ Lett.}}
  \textbf{\bibinfo{volume}{59}}, \bibinfo{pages}{381--384}
  (\bibinfo{year}{1987}).

\bibitem{Harris1989}
\bibinfo{author}{Harris, T.~E.}
\newblock \emph{\bibinfo{title}{{The theory of branching processes}}}
  (\bibinfo{publisher}{Dover}, \bibinfo{address}{New York},
  \bibinfo{year}{1989}).

\end{thebibliography}

\begin{addendum}
% \item[Supplementary Information] is linked to the online version of the
% paper at www.nature.com/nature.
 \item[Acknowledgements] 
This work was supported by Grants-in-Aid from the Ministry of Education, Science, Sports, and
 Culture of Japan: Grant numbers 16200025, 17022020, 17650100, 18019019,
 18047014, and 18300079.
 \item[Author Information] 
%Reprints and permissions information is available at npg.nature.com/reprintsandpermissions.
The authors declare that they have no
competing financial interests.
Correspondence and requests for materials should be addressed to ttakuma@mbs.med.kyoto-u.ac.jp.
\end{addendum}

%Kaneko 16200025, 17022020,17650100
%Aoyagi 18047014, 18019019, 18300079

%\begin{document}

\end{document}